\documentclass[12pt]{article}
\usepackage{amsmath,amssymb}
\usepackage{graphicx}
\usepackage{amsmath}
\usepackage{float}
\usepackage[toc,page]{appendix}
\usepackage{xcolor}

\newcommand {\be}{\begin{equation}}
 \newcommand {\ee}{\end{equation}}
 \newcommand {\bea}{\begin{array}}
 
 \newcommand {\eea}{\end{array}}

\textwidth=6.5in \hoffset=-.75in \voffset=-1in \numberwithin{equation}{section}
\numberwithin{figure}{section}

\def\e{{\epsilon}}

\def\l{\ell}

 \def\p{\partial}

\def\0{{(0)}}
\def\1{{(1)}}
\def\2{{(2)}}
 \def\cL{{\cal L}}

\def\<{\langle }
\def\>{\rangle }
\def\[{\left[}
\def\]{\right]}

\def\z{{\zeta}}

\begin{document}

\begin{titlepage}

\vskip1cm
\begin{center}
{~\\[140pt]{ \LARGE {{Holographically description of  Kerr-Bolt \\ Black hole in terms of the warped conformal field theory }}}\\[-20pt]}
\vskip2cm

\end{center}
\begin{center}
	{M. R. Setare \footnote{E-mail: rezakord@ipm.ir}\hspace{1mm} ,
		A. Jalali \footnote{E-mail: alijalali.alijalali@gmail.com}\hspace{1.5mm} \\
		{\small {\em  {Department of Science,\\
					Campus of Bijar, University of Kurdistan, Bijar, Iran
					}}}}\\
\end{center}
\begin{abstract}
Recently it has been speculated that a set of   diffeomorphisms exist which  act non-trivially on the horizon of some black holes such as Kerr and Kerr-Newman black hole. 
Using this symmetry in  covariant phase space  formalism one can obtain conserved charges as surface integrals on
the horizon. Kerr-Bolt spacetime is well-known for its asymptotic topology and has been studied widely in recent years. In this work we are going  to   study Kerr-Bolt black hole and provide  a holographic description of it in term of warped conformal field theory.
\end{abstract}
\vspace{1cm}


\end{titlepage}

\section{Introduction}
Not many decades ago, Vafa and Strominger have derived the microscopic entropy of five-dimensional
extremal BPS black holes in the context of string theory \cite{Strominger:1996sh,Horowitz:1996fn}. It soon became clear that the string theory was not the key feature as well as the fact that  the conformal symmetry plays a central role in their achievement. In fact, AdS factor which appears in the near horizon geometry is the reason of conformal symmetry, and then AdS/CFT help us to understand this agreement.

The story has got even more interesting when Castro et. al \cite{Castro:2010fd} (see also \cite{3}) introduced hidden conformal symmetry of black hole.
They have   expanded the notion of explicit conformal invariance associated to the near horizon AdS$_3$ geometry to the hidden one on the Kerr background.
In fact Kerr black hole with mass $M$ and spin $J$ has a hidden conformal symmetry which is made manifest when considering near region in the phase space rather than spacetime\cite{Castro:2010fd}. It is also worth pointing out that the hidden conformal symmetry of many others balck hole has been studied widely in recent years   \cite{Chen:2010as,Wang:2010qv,Chen:2010bd,Setare:2010cj}.

On the other hand, two sets of Virasoro diffeomorphisms acting non trivially on the Kerr horizon  was introduced  recently \cite{Haco:2018ske}. Covariant phase space was used there to produce the central charge   $c_L=c_R=12 J$.
Using  the Cardy formula in a CFT$_2$ as
$$S_{\text{Cardy}}=\frac{\pi^2}{3}\left(c_LT_L+c_RT_R\right)$$
along with right-moving and left-moving temperature in \cite{Castro:2010fd},  authors in  \cite{Haco:2018ske} (see also \cite{Haco:2019ggi}) could interestingly reproduce Bekenstein-Hawking area law of the Kerr black hole.
Seeking an alternative description of Kerr black hole, authors in \cite{Aggarwal:2019iay} have suggested warp conformal symmetry. They examine this significant  question of whether the traditional AdS/CFT description of Kerr black hole is unique? The answer is no.

In recent years, another holographic plan has been in progress  in which the geometries is no longer a local $SL(2,R)\times SL(2,R)$ symmetry and also its dual field theory is not CFT$_2$. This plan provides a new type of field theories which is called warped conformal field theory (WCFT) \cite{Hofman:2011zj,Detournay:2012pc}.
Generally, in contrast to CFTs, which have both independent right and left  scaling symmetry in the light cone coordinate, i.e.
\begin{align*}
x^-\to \lambda x^-\nonumber\\
x^+\to \lambda x^+
\end{align*}
WCFTs are required to be invariant only under left scaling translation.
These symmetries were studied in \cite{Detournay:2012pc} and reveal
  the following conclusions. A $2d$ translation-invariant theory with a chiral scaling symmetry
must have an extended local algebra. There are two minimal options for this algebra. One
is the usual CFT case with two copies of the Virasoro algebra. The other possibility is one
Virasoro algebra plus a U(1) Kac-Moody algebra. The last one is called a Warped Conformal
Field Theory \cite{Detournay:2012pc}.

Kerr black hole, apart from usual CFT$_2$ description, could be described in term of WCFT and reproduce Bekenstein-Hawking entropy. In the following work, we are going to study behavior of the Kerr-Bolt black hole, which is well-known for its asymptotically topology and has been studied
widely in recent years \cite{Clarkson:2003wa},  in term of WCFT.\\
We know that there is a hidden conformal symmetry of the Klein-Gordon
operator on the Kerr-Bolt background \cite{Chen:2010bd,Setare:2010cj}, which is made manifest when consider the near region of phase space instead of spacetime. In the other hand we have an explicit identification of two
sets of Virasoro diffeomorphisms acting on the horizon of Kerr-Bolt black hole \cite{Setare:2019fxa}. In this paper we show that the arguments used in \cite{Chen:2010bd,Setare:2010cj,Setare:2019fxa} which describe the Kerr-Bolt black holes
 in terms of a CFT, are not unique. We show that there is a description of Kerr-Bolt black holes in terms of WCFT also.
These investigation in the back ground of Kerr black hole have been done respectively in \cite{Castro:2010fd}, \cite{Haco:2018ske}, \cite{Aggarwal:2019iay}. In \cite{Castro:2010fd} it has been shown that hidden conformal symmetry of the Klein-Gordon
operator in "near region" phase space can be described by a CFT$_{2}$. In \cite{Haco:2018ske} using the phase space formalism the authors have found a near-horizon Virasoro algebra. In the other hand the authors of \cite{Aggarwal:2019iay} have shown that a description of Kerr black holes in terms
of a WCFT can appear as natural as a CFT$_{2}$ one. \footnote{In \cite{Aggarwal:2019iay} it has also been shown that how a mild modification to the "near region" allows to obtain a hidden warped conformal symmetry for the Klein-Gordon
operator on the Kerr background. Here in our Kerr-Bolt black holes case also we expect such warped hidden symmetry can appear by mild modification to the   "near region" phase space.}  \\
 Our study is organized as following sections: In the next section we take a brief look at the warped conformal symmetry. In the section 3 Kerr-Bolt spacetime will be introduced
  briefly as well as its hidden conformal symmetry. In section 4 we will expand Kerr-Bolt metric in the conformal coordinate and look into the its warped symmetry. In this section we also use the covariant phase space formalism to calculate conserved charge associated to the warped symmetry and we will show our proposal about holographic description of Kerr-Bolt black hole in term of WCFT. In section 5 we apply the Cardy entropy formula and reproduce the Bekenstein-Hawking entropy. Last section includes our concluding remarks.

\section{Warped conformal symmetry}
There are at least four global symmetries associated to the 2D Poincare and scale invariant
quantum field theory which act on the light-cone coordinate $(x^\pm=x\pm t)$ as
\begin{align}
&x^-\to x^-+a\,,\qquad x^+\to x^++b\nonumber\\
&x^-\to \lambda^- x^-\,,\qquad  x^+\to \lambda^+ x^+
\end{align}
If unitarity and discrete  spectrum of the dilation operator for
$\lambda^+=\lambda^-$ add to the above transformations, it was shown in \cite{Polchinski:1987dy} that
the four global symmetries are enhanced to left and
right infinite-dimensional conformal symmetries. Warped conformal field theory is arisen form the special case in which there are three global symmetries: two translational  and a chiral
“left” dilational. Authors in \cite{Hofman:2011zj} studied these symmetries and shown adding  locality, unitary and a discrete
non-negative dilational spectrum but not Lorentz invariance are sufficient to conclude that at least  there are two infinite-dimensional
sets of local symmetries.

Now consider the two operators as
\begin{align}
L_n=-\frac{i}{2\pi}\int dx^-\zeta(x^-)T(x^-)\,,\qquad
P_n=-\frac{1}{2\pi}\int dx^-\xi(x^-)P(x^-),
\end{align}
where $T(x^-)$ is generator of infinitesimal coordinate transformation in $x^-$ and $P(x^-)$ is generator
of $x^-$ -dependent infinitesimal coordinate translations in $x^+$.
Choosing  functions $\zeta_n=(x^-)^{n+1}$ and
$\xi_n=(x^-)^{n}$
it could be shown that $L_{-1}$ is the left moving translation while $P_n$ parameterize the $U(1)$ Kac-Moody current algebra \cite{Hofman:2011zj}.
The algebra and  $L_n$ and $P_n$ commutation relations obey a Virasoro-Kac-Moody algebra:
\begin{align}
[L_n,L_m]&=(m-n)L_{m+n}+\frac{c}{12}n(n^2-1)\delta{m+n}\nonumber\\
[L_n,P_m]&=m P_{n+m}\\
[P_n,P_m]&=k\frac{n}{2}\delta_{m+n}
\end{align}
where  $c$ is central charge and $k$ is $U(1)$ level. These generators both can be used to provide entropy via WCFT partition  function. See \cite{Aggarwal:2019iay} for more detail on this subject.
\section{Hidden conformal symmetry of the Kerr-Bolt black hole}
We begin with Kerr-Bolt geometry.
The Kerr-Bolt spacetime with NUT charge $p$ and rotational parameter $a$, is given by the line element
\begin{eqnarray}
ds^{2}&=&-\frac{\Delta(r)}{\rho^{2}}[dt+(2p\cos(\theta)-a\sin^{2}(\theta)) d\phi]^{2}
+\frac{\sin^{2}(\theta)}{\rho^{2}}[a dt-(r^{2}+p^{2}+a^{2})d\phi]^{2}
\nonumber \\
&+& \frac{\rho^{2}dr^{2}}{\Delta(r)}+\rho^{2} (d\theta)^{2},
\label{intro1}
\end{eqnarray}
where
\begin{eqnarray}
\rho^{2}&=&r^{2}+(p+a\cos(\theta))^{2},\\
\label{intro2}
\Delta(r)&=& r^2-2 M r +a^{2}-p^{2}.
\label{intro3}
\end{eqnarray}
The Kerr-Bolt spacetime (\ref{intro1}) is exact solution to Einstein equations. The inner $r_-$ and outer $r_+$ horizons of spacetime (\ref{intro1}) are $r_{-}=M-\sqrt{M^2+p^2-a^2}$ and $r_+=M+\sqrt{M^2+p^2-a^2}$.

To reveal the hidden conformal symmetry of the Kerr-Bolt metric, we consider massless scalar field in the Kerr-Bolt background and then focus on its near region behavior.
The Klein-Gordon (KG) equation for the massless scalar field $\Phi$
\begin{equation}
\Box\Phi=\frac{1}{\sqrt{-g}}\partial_{\mu}(g^{\mu\nu}\partial_{\nu})\Phi=0,
\label{intro4}
\end{equation}
can be separated  as
\begin{equation}
\Phi(t,r,\theta,\phi)=\exp(-i m\phi+i\omega t) S(\theta) R(r).
\label{intro5}
\end{equation}
Then, the equation (\ref{intro4}) reduces to
\begin{equation}
[\partial_{r}(\Delta(r)\partial_{r})+\frac{1}{\sin\theta} \partial_{\theta}(\sin\theta\partial_{\theta})-\frac{(C(\theta)\omega+m)^{2}}{\sin^{2}\theta}+\frac{(D(r)\omega-m a)^{2}}{\Delta(r)}] S(\theta) R(r)=0,
\label{intro6}
\end{equation}
where the functions $C(\theta)$ and $D(r)$ are given by
\begin{eqnarray}
C(\theta)&=&2p\cos(\theta)-a\sin^{2}(\theta),\\
D(r)&=&r^{2}+p^{2}+a^{2}.
\label{intro7}
\end{eqnarray}
The separated equation for $S(\theta)$ is given by
\begin{equation}
[\frac{1}{\sin(\theta)} \partial_{\theta}(\sin(\theta)\partial_{\theta})+f(\theta)] S(\theta)=-\lambda S(\theta),\label{intro8}
\end{equation}
where
\begin{equation}
f(\theta)=\frac{-4n^2\omega^2-m^2-4mn\omega\cos(\theta)}{\sin^{2}(\theta)}+a^2\omega^2\cos^2(\theta)+2na\omega^2\cos(\theta)~
\label{intro9}
\end{equation}
and $\lambda$ is the separation constant.
Considering low frequency scalar field and in the near region of geometry i.e.
\begin{eqnarray}
\omega M, \omega p  &\ll& 1, \\
r &\ll& \frac{1}{\omega},
\label{intro10}
\end{eqnarray}
the angular equation (\ref{intro8}) significantly simplifies as \cite{Setare:2010cj,Castro:2010fd}
\begin{equation}
[\frac{1}{\sin(\theta)} \partial_{\theta}(\sin(\theta)\partial_{\theta})-\frac{m^2}{sin^2\theta}] S(\theta)=-l(l+1)S(\theta).\label{intro88}
\end{equation}
as well as the radial part of KG equation becomes
\be \begin{array}{cc}
	[\partial_{r}(\Delta\partial_{r})+\frac{-4Mmar\omega+m^{2}a^{2}-4n^{2}ma\omega}{\Delta(r)}+(\Delta+4(Mr+n^{2})-a+4n)
	\omega^{2}]R(r)=l(l+1)R(r),
\end{array} \label{intro11}\ee
Conformal coordinate come in handy to show $SL(2,R)\times SL(2,R)$ symmetry of the near-region
scalar field equation in the Kerr-Bolt background more explicit. Confromal coordinate ($\omega^\pm,\ y$) in the Kerr-Bolt space time defines as
\begin{eqnarray}
\omega^{+}&=&\sqrt{\frac{r-r_{+}}{r-r_{-}}}\exp(2\pi T_{R}\phi+2n_{R}t),
\label{intro14}\\
\omega^{-}&=&\sqrt{\frac{r-r_{+}}{r-r_{-}}}\exp(2\pi T_{L}\phi+2n_{L}t),
\label{intro15}\\
y&=&\sqrt{\frac{r_{+}-r_{-}}{r-r_{-}}}\exp(\pi( T_{R}+T_{L})\phi+(n_{R}+n_{L})t),
\label{intro16}
\end{eqnarray}
where
\begin{equation}
T_{R}=\frac{r_{+}-r_{-}}{4\pi a}, ~~~~~~T_{L}=\frac{r_{+}+r_{-}}{4\pi a}+\frac{p^{2}}{2\pi a M},
\label{intro17}
\end{equation}
and $n_{R}=0,n_{L}=-\frac{1}{4M}$. Now we define right and left moving vector field as
\begin{equation}
	H_1=i\partial_{+},~~~~H_0=i(\omega^{+}\partial_{+}+\frac{1}{2}y
	\partial_{y}),~~~~~H_{-1}=i((\omega^{+})^2\partial_{+}+\omega^{+}y\partial_{y}-y^2\partial_{-}),~~~~
	\label{intro19}
\end{equation}
and
\begin{equation}
	\bar{H}_1=i\partial_{-},~~~~\bar{H}_0=i(\omega^{-}\partial_{-}+\frac{1}{2}y\partial_{y}),
	~~~~~\bar{H}_{-1}=i((\omega^{-})^2\partial_{-}+\omega^{-}y\partial_{y}-y^2\partial_{+}),
	\label{intro20}
\end{equation}
respectively. Some algebraic calculations show that
the vector fields (\ref{intro19}) satisfy the $SL(2,R)$ algebra \cite{Setare:2010cj}
\begin{equation}
~~[H_0,H_{\pm1}]=\mp i H_{\pm 1},~~~~~~~~[H_{-1},H_1]=-2iH_0,~~
\label{intro21}
\end{equation}
and similarly for $\bar{H}_1,\bar{H}_0$ and $\bar{H}_{-1}$.
The Casimir operator $H^2=-H_{0}^2+\frac{1}{2}(H_1H_{-1}+H_{-1}H_{1})$ (as well as the other Casimir operator $\bar{H}^2$) reduces to the radial equation (\ref{intro11}),
\be \begin{array}{cc}
	H^2R(r)=\bar{H}^2R(r)=l(l+1)R(r)~~~
\end{array} \label{intro29}\ee
It is important to   note  that above symmetry is only a local $SL(2,R)_L \times SL(2,R)_R$ hidden conformal symmetry.
To be more accurate, vectors in \eqref{intro21}
which generate the SL(2,R) symmetries are not  periodic under $\phi \sim \phi +2\pi$ identification and so are not globally defined \cite{Castro:2010fd,Setare:2010cj}.
\section{Warped symmetries in the Kerr-Bolt geometry}
Conserved charge associated to the certain symmetry has   long story in the  general relativity. Wald and his team have made many efforts to clarify the matters \cite{Compere:2018aar,wz,iw}. They have introduced the concept of covariant phase space by which one can evaluate the conserved charge associated to any symmetry in the gravitational theories. We will introduce a  set of diffeomorphisms which act in the near horizon geometry of the Kerr-Bolt black hole, and study the linearized charges associated to them, by using Wald method.
Consider the following vectors
\begin{align} \label{virU1}
\zeta(\epsilon)&= \epsilon(w^+)\, \partial_++{1\over 2}\partial_+\epsilon(w^+) \,y\partial_y~,\cr
p(\hat\e)&=  ~\hat\e(w^+)\, (w^- \p_- +{y\over 2} \p_y) ~ ,
\end{align}
where $\epsilon$ and $\hat \e$ are arbitrary functions of $w^+$, which restrict   such that the vector fields (\ref{virU1}) are periodic under
$\phi\sim \phi+2\pi$.
A complete
set of such functions is
 \be\label{csf}
\e_n={2 \pi T_R}(w^+)^{1+{in \over 2 \pi T_R}}~,\qquad \hat\e_{n'}=(w^+)^{{in' \over 2 \pi T_R}}~,
\ee
It is easy to show that above vectors satisfy following communication relations:
\begin{align} \label{VKM}
i [\zeta_m, \zeta_n ]&= (m-n) \zeta_{m+n}~, \cr
i [\zeta_m, p_n]&= n\, p_{m+n}~, \cr
i [p_m, p_n]&= 0~,
\end{align}
which is a Virasoro-Kac-Moody (VKM) algebra without any central charge.

It must be  indicated that although VKM algebra can be seen manifestly in the phase space formalism in terms of $\zeta$ vectors,  the vectors in \eqref{intro19} and  \eqref{intro20} obey the  $SL(2,R)$  algebra, instead of the VKM algebra. In fact, as authers in \cite{Aggarwal:2019iay} have discussed, we have not considerd the $H$ vectors in \eqref{intro19} and  \eqref{intro20} as the basis for $CFT_2$ energy eigenstates. Instead, according to the
\begin{equation}
	-i(2\pi T_R){H}_0=(2\pi T_R)(\omega^{+}\partial_{+}+\frac{1}{2}y\partial_{y})=\zeta_0(\epsilon),
	~~~~~,	-i\bar{H}_0=(\omega^{-}\partial_{-}+\frac{1}{2}y\partial_{y})=P_0(\epsilon),
	\label{neweq}
\end{equation}
relations, we tried to interpret them as the WCFT zero mode symmetry vectors.\\
 To leading and subleading order around
 the bifurcation surface, the Kerr-Bolt metric in \eqref{intro1} becomes
 \begin{eqnarray}
 \label{kbm}
 \begin{split} ds^2&= {4 \rho_+^2 \over y^2} d w^+ dw^-
 +  {16 J^2 \sin^2\theta \over y^2 \rho_+^2} dy^2 +\rho_+^2 d\theta^2 \\[6pt]  &
 - {2w^+ (8\pi J)^2 T_R(T_R+T_L) +p(64 \pi a T_RJ\cos\theta) \over y^3 \rho_+^2} dw^- dy \\[6pt]  &
 + {8 w^- \over y^3 \rho_+^2} \big(A+B\big) dw^+ dy \\[6pt] &
 + \cdots, \end{split}
 \end{eqnarray}
 where
 \begin{eqnarray}
 A&=&- (4\pi J)^2T_L(T_R+T_L) + (4 J^2 + 4\pi J a^2 (T_R+T_L)  + a^2 \rho_+^2) \sin^2\theta\nonumber \\
 B&=&-4pa\cos\theta\big(m^2+p^2+1/2 \rho_+^2\big)
 \end{eqnarray}
and  here  $\rho_+$  is given by
 \begin{equation}
 \rho_+^2=r_+^{2}+(p+a\cos(\theta))^{2}=
 (m+\sqrt{-a^2+m^2+p^2})^{2}+(p+a\cos(\theta))^{2}\\
 \end{equation}
 The conserved charge associated to the VKM algebra as well as  any others has two parts, the first refers to the Iyer-Wald charge, i.e. $\delta Q_{IW}$, while the second is Wald-Zoupas charge, i.e. $\delta Q_{WZ}$. So the conserved charge reads as:
 \begin{align}
 \delta{\cal{Q}}(\zeta,h,g)&=\frac{1}{16\pi}\left(\delta{\cal{Q}}_{IW}(\zeta,h,g)+\delta{\cal{Q}}_X(\zeta,h,g)\right)\nonumber\\
 &=\frac{1}{16\pi}\int_{\Sigma_{\text{bif}}}\ast F_{IW}+\frac{1}{16\pi}\int_{\Sigma_{\text{bif}}}\zeta\cdot(\ast X)
 \end{align}
 where
 \begin{align}
 \label{fab}
 \begin{split} F_{IWab} = \frac{1}{2}\nabla_a\zeta_bh
 +\nabla_ah^c{}_b \zeta_c
 +\nabla_c\zeta_a\ h^c{}_b
 +\nabla_ch^c{}_a\ \zeta_b
 -\nabla_ah\ \zeta_b - a \leftrightarrow b. \end{split}
 \end{align}
 and X is a  one-form constructed from the geometry of the spacetime as follow \cite{Haco:2018ske}
 \begin{equation}
 \label{counter}  X = 2dx^ah_a^{~b} \Omega_b\,.
 \end{equation}
 where
 \begin{equation}
 \Omega_a=q_a^c n^b\nabla_c \l_b\label{ox1}\,.
 \end{equation}
 Two vectors $\l^a$ and $n^a$ are normal to the future and past horizon respectively and have been chosen so that $\l\cdot n=-1$. We also consider $q_{ab}=g_{ab}+\l_an_b+n_a\l_b$ as the induced metric on the bifurcation surface \cite{Haco:2018ske}. It must be noted that all above integrals are taken over bifurcation surface.

 Assuming integrability, the conserved charge associated to   these symmetries also form an algebra under Dirac bracket:
 \begin{equation}
 \{  {\cal{Q}}_n,  {\cal{Q}}_m\}=(m-n) {\cal{Q}}_{m+n}+K_{m,n},
 \end{equation}
 where the central extension is given by \cite{Compere:2018aar}
 \begin{equation}
 \label{cterm} K_{m,n}=\delta {\cal{Q}}(\z_n,\cL_{\z_m}g;g).
 \end{equation}
There are three different central  charges that related to the KMV algebra. First of all is the Virasoro charge. We have shown in the previous work \cite{Setare:2019fxa} that Virasoro generator obeys the following algebra:
\be \label{cterm}
[L_n, L_m]= (m-n) L_{m+n} +  K_{m,n}~, \qquad K_{m,n}=\delta \mathcal{Q}(\z_n,\mathcal L_{\z_m}g;g)~,
\ee
 where
 \begin{align}\label{eq:k1}
 \delta \mathcal{Q}(\zeta_n,\mathcal L_{\zeta_m}g;g)&=\delta \mathcal{Q}_{\text{IW}}(\zeta_n,\mathcal L_{\zeta_m}g;g)+\delta \mathcal{Q}_{\text{WZ}}(\zeta_n,\mathcal L_{\zeta_m}g;g)\nonumber\\
 \delta \mathcal{Q}_{\text{IW}}(\zeta_n,\mathcal L_{\zeta_m}g;g)&= 2J {T_R\over T_L+T_R-\alpha}\left((2\pi T_R)^2m + m^3\right)\delta_{n+m}~,\\\nonumber \delta\mathcal{Q}_{\text{WZ}}(\zeta_n,\mathcal L_{\zeta_m}g;g)&= J {T_L-T_R-\alpha\over T_L+T_R-\alpha}\left((2\pi T_R)^2m + m^3\right)\delta_{n+m}~,
 \end{align}
 where $\alpha$ is the constant and defines by $\alpha=\frac{p^2}{2\pi M a}$.
 Adding up the above two parts, we obtain the following central charge associated to the Virasoro algebra:
 \be\label{eq:ctot}
 K_{m,n}= J\,m^3\, \delta_{n,-m} ~,
 \ee
 Another conserved charge is the Kac-Moody conserved charge which turns out as
 \be\label{eq:kmterm}
 [P_n, P_m]=  k_{m,n}~, \qquad k_{m,n}=\delta \mathcal{Q}(p_n,\mathcal L_{p_m}g;g)~.
 \ee
 where
  \begin{align}\label{eq:k1}
 \delta \mathcal{Q}(p_n,\mathcal L_{p_m}g;g)=\delta \mathcal{Q}_{\text{IW}}(p_n,\mathcal L_{p_m}g;g)+\delta \mathcal{Q}_{\text{WZ}}(p_n,\mathcal L_{p_m}g;g)
 \end{align}
 Carrying out some calculations, we find out that the only nonzero component of $F^{ab}$ tensor in \eqref{fab} that contributes to the $\delta \mathcal{Q}_{\text{IW}}$ charge is $F^{-y}$ so
  \begin{align}
\delta \mathcal{Q}_{\text{IW}}(p_n,\mathcal L_{p_m}g;g)=\frac{1}{16\pi}\int_{\Sigma_{\text{bif}}}\ast F_{IW}=- 2J {T_L-\alpha\over T_L+T_R-\alpha} m\,\delta_{n+m}\,.
 \end{align}
 Follow the same steps as mentioned in  \cite{Haco:2018ske,Setare:2019fxa} we get the following result:
  \begin{align}
 \delta \mathcal{Q}_{\text{WZ}}(p_n,\mathcal L_{p_m}g;g)=\frac{1}{16\pi}\int_{\Sigma_{\text{bif}}}\zeta\cdot(\ast X)= J {T_L-T_R-\alpha\over T_L+T_R-\alpha}m\,\delta_{n+m}\,,
 \end{align}
 and then
 \be
 k_{m,n}=-J\, m\,\delta_{n,-m}~.
 \ee
 Finally, we analyze the mixed central charge for the KMV algebra in the Kerr-Bolt background and find out
 \begin{align}
 \delta \mathcal{Q}_{\rm IW}(\zeta_n,\mathcal L_{p_m}g;g)=i J {T_R-T_L-\alpha\over T_L+T_R-\alpha}\bigg(i m ({2\pi T_R)}+m^2\bigg)\,\delta_{n+m} ~,\cr
 \delta \mathcal{Q}_{\rm WZ}(\zeta_n,\mathcal L_{p_m}g;g)=-i J {T_R-T_L-\alpha\over T_L+T_R-\alpha}\bigg(i m ({2\pi T_R)}+m^2\bigg)\,\delta_{n+m} ~,
 \end{align}
which cancel each other and hence
 \begin{align}
 \mathfrak{K}_{m,n}=0
  \end{align}
  At a glance, our achievements could be listed as:
  \begin{align}
  [L_n, L_m]&= (m-n) L_{m+n} + \frac{c}{12}m^3\delta_{n,-m}~,\cr
  [L_n, P_m]&= m P_{n+m}~,\cr
  [P_n, P_m]&=k \frac{m}{2}\delta_{n,-m}~,
  \end{align}
  with
  \be \label{CentralExtensions}
  c =12 J ~,\qquad k=-2J~.
  \ee
 which explicitly shows  central charges associated to the Virasoro-Kac-Moody algebra depend on the angular momentum of the Kerr-Bolt black hole.
    This results also support the our proposal that the Kerr-Bolt black hole could be described holographically in terms of a warped conformal field theory.
 \section{Black hole entropy from warped symmetries}
 In this part we want use the warped version of Cardy formula to reproduce the Bekenstein-Hawking entropy of Kerr-Bolt black hole. According to the Cardy
 formula for warped CFT, the entropy is given by
  \begin{align}
  S_{WCFT}=4\pi^{2}i(T_{L}+T_{R})P_{0}^{vac}+8\pi^{2}T_{R}L_{0}^{vac}
  \end{align}
  where $P_{0}^{vac}$, $L_{0}^{vac}$ are the vacuum values of the zero mode in WCFT. $P_{0}^{vac}$,and $L_{0}^{vac}$ are related to each other by following relation
  \begin{align}
  L_{0}^{vac}=\frac{-c}{24}+\frac{(P_{0}^{vac})^{2}}{k}.
  \end{align}
  There is an ambiguity in the value of $P_{0}^{vac}$, in order to fixe this ambiguity, we should consider
  \begin{align}
  L_{0}^{vac}=0.
  \end{align}
   So, by inserting the above equation in Eq.(5.2), and using the values of $c=12J$, $k=-2J$, we obtain
   \begin{align}
  P_{0}^{vac}=-J^{2}.
  \end{align}
   By substituting $T_{L}$, $T_{R}$ from Eq.(3.18) and above $P_{0}^{vac}$, into Eq. (5.1), we obtain following expression for entropy of WCFT
   \begin{align}
  S_{WCFT}=\frac{2\pi|J|}{a}(r_{+}+\frac{p^{2}}{M})=2\pi M(r_{+}+\frac{p^{2}}{M}),
  \end{align}
 which is in complete agreement with following the Bekenstein-Hawking entropy formula \cite{mann}
  \begin{align}
  S_{BH}=\pi(r_{+}^{2}+a^{2}+p^{2}),
  \end{align}
  upon substitution $r_{+}=M+\sqrt{M^{2}+p^{2}-a^{2}}$.
 \section{Discussion}
In this paper we have shown that warped conformal field theory could be a holographic description of the Kerr-Bolt black
hole. Previously we have considered the four-dimensional spacetimes with rotational parameter and NUT
twist and have shown they have a hidden conformal symmetry \cite{set}. The spacetimes with NUT twist
have been studied extensively in regard to their conserved charges, maximal mass conjecture
and D-bound in \cite{Clarkson:2003wa}.
After a brief look
at the warped conformal symmetry, we have introduced Kerr-Bolt spacetime  as well as its hidden conformal symmetry. Then
we have expanded Kerr-Bolt metric in the conformal coordinate and have looked into the its warped
symmetry. We have used the covariant phase space formalism to calculate
conserved charge associated to the warped symmetry and we have shown our proposal about
holographic description of Kerr-Bolt black hole in term of warped conformal field theory.

\section{Acknowledgements}
 We thank the anonymous referee for important
comments.



\begin{thebibliography}{99}
	\bibitem{Strominger:1996sh}
	A.~Strominger and C.~Vafa,
	``Microscopic origin of the Bekenstein-Hawking entropy,''
	Phys.\ Lett.\ B {\bf 379}, 99 (1996)
	[hep-th/9601029].
	\bibitem{Horowitz:1996fn}
	G.~T.~Horowitz and A.~Strominger,
	``Counting states of near extremal black holes,''
	Phys.\ Rev.\ Lett.\  {\bf 77}, 2368 (1996).
	\bibitem{Castro:2010fd}
	A.~Castro, A.~Maloney and A.~Strominger,
	Phys.\ Rev.\ D {\bf 82}, 024008 (2010),
		[arXiv:1004.0996 [hep-th]].
\bibitem{3}M. Guica, T. Hartman, W. Song, A. Strominger, Phys. Rev. D. 80, 124008, (2009), arXiv:0809.4266 [hep-th].
	 \bibitem{Chen:2010as}
	C.~M.~Chen and J.~R.~Sun,
	``Hidden Conformal Symmetry of the Reissner-Nordstrom Black Holes,''
	JHEP {\bf 1008}, 034 (2010)
	[arXiv:1004.3963 [hep-th]].
	 \bibitem{Wang:2010qv}
	Y.~Q.~Wang and Y.~X.~Liu,
	``Hidden Conformal Symmetry of the Kerr-Newman Black Hole,''
	JHEP {\bf 1008}, 087 (2010)
	[arXiv:1004.4661 [hep-th]].
	\bibitem{Chen:2010bd}
	B.~Chen, A.~M.~Ghezelbash, V.~Kamali and M.~R.~Setare,
	``Holographic description of Kerr-Bolt-AdS-dS Spacetimes,''
	Nucl.\ Phys.\ B {\bf 848}, 108 (2011)
	[arXiv:1009.1497 [hep-th]].
	\bibitem{Setare:2010cj}
	M.~R.~Setare and V.~Kamali,
	``Hidden Conformal Symmetry of Extremal Kerr-Bolt Spacetimes,''
	JHEP {\bf 1010}, 074 (2010)
	[arXiv:1011.0809 [hep-th]].
	  \bibitem{Haco:2018ske}
	  S.~Haco, S.~W.~Hawking, M.~J.~Perry and A.~Strominger,
	  ``Black Hole Entropy and Soft Hair,''
	  arXiv:1810.01847 [hep-th].
\bibitem{Haco:2019ggi}
  S.~Haco, M.~J.~Perry and A.~Strominger,
  ``Kerr-Newman Black Hole Entropy and Soft Hair,''
  arXiv:1902.02247 [hep-th].
	\bibitem{Aggarwal:2019iay}
	A.~Aggarwal, A.~Castro and S.~Detournay,
	``Warped Symmetries of the Kerr Black Hole,''
	arXiv:1909.03137 [hep-th].
	\bibitem{Hofman:2011zj}
	D.~M.~Hofman and A.~Strominger,
	Phys.\ Rev.\ Lett.\  {\bf 107}, 161601 (2011),
		[arXiv:1107.2917 [hep-th]].
	\bibitem{Detournay:2012pc}
	S.~Detournay, T.~Hartman and D.~M.~Hofman,
	Phys.\ Rev.\ D {\bf 86}, 124018 (2012),
		[arXiv:1210.0539 [hep-th]].
	\bibitem{Clarkson:2003wa}
	R.~Clarkson, A.~M.~Ghezelbash and R.~B.~Mann,
	Phys.\ Rev.\ Lett.\  {\bf 91}, 061301 (2003),
		[hep-th/0304097].
\bibitem{Setare:2019fxa}
  M.~R.~Setare and A.~Jalali, Int. J. Mod. Phys. A.35, 25, 2050156 (2020).
  arXiv:1911.02550 [hep-th].
	\bibitem{Polchinski:1987dy}
	J.~Polchinski,
	Nucl.\ Phys.\ B {\bf 303}, 226 (1988).
		
	  \bibitem{Compere:2018aar}
	A.~Fiorucci and G.~Comp\`ere,
	``Advanced Lectures in General Relativity,''
	arXiv:1801.07064 [hep-th].
	\bibitem{wz}
	R.~M.~Wald and A.~Zoupas,
	``A General definition of 'conserved quantities' in general relativity and other theories of gravity,''
	Phys.\ Rev.\ D {\bf 61}, 084027 (2000)
	[gr-qc/9911095].
	\bibitem{iw} V. Iyer and R.M.Wald, Phys. Rev. {\bf D50}, 846 (1994).


















\bibitem{mann}
 R. Mann, Phys. Rev. D61, 084013 (2000).

\bibitem{set}A. M. Ghezelbash, V. Kamali, M. R. Setare,  Phys. Rev. D82, 124051, (2010), arXiv:1008.2189 [hep-th].


\end{thebibliography}
\end{document}